\documentclass[a4paper]{article}
\usepackage[affil-it]{authblk}

\usepackage{hyperref}
\usepackage{booktabs}
\usepackage{braket}
\usepackage{enumitem}
\usepackage{amssymb}
\usepackage{amsmath}
\usepackage{graphicx}
\usepackage{verbatim}
\usepackage[ruled,noend]{algorithm2e}
\usepackage{float}
\usepackage{mathtools}

\providecommand{\keywords}[1]
{
	\small	
	\textbf{\textit{Keywords---}} #1
}

\begin{document}
	
	\title{Two Efficient Measurement Device Independent Quantum Dialogue Protocols}
	
	\author{Nayana Das%
		\thanks{Email address: \texttt{dasnayana92@gmail.com} }}
	\affil{Applied Statistics Unit, Indian Statistical Institute, India}
	
	\author{Goutam Paul %
		\thanks{Email address: \texttt{goutam.paul@isical.ac.in}}}
	\affil{Cryptology and Security Research Unit, R. C. Bose Centre for Cryptology and Security, Indian Statistical Institute, Kolkata, India.\\}
	
	\date{}
	
	\maketitle
	\begin{abstract}
		Quantum dialogue is a process of two way secure and simultaneous communication using a single channel. Recently, a Measurement Device Independent Quantum Dialogue (MDI-QD) protocol has been proposed (Quantum Information Processing 16.12 (2017): 305). To make the protocol secure against information leakage, the authors have discarded almost half of the qubits remaining after the error estimation phase. In this paper, we propose two modified versions of the MDI-QD protocol such that the number of discarded qubits is reduced to almost one-fourth of the remaining qubits after the error estimation phase. We use almost half of their discarded qubits along with their used qubits to make our protocol more efficient in qubits count. We show that both of our protocols are secure under the same adversarial model given in MDI-QD protocol.
	\end{abstract}
	
	
	\section{Introduction}
	\label{intro}
	Quantum cryptography uses the unconventional properties of quantum mechanics like entanglement theory \cite{Einstein1935}, no cloning theorem \cite{Wootters1982} etc. to perform  cryptographic tasks. It provides unconditional security and innovative ways of communicating. There  are  many  modes of quantum communication, such as Quantum Key Distribution (QKD) \cite{bb84,Ekert91,Ben92,Shor00,Long02,Hwang03,lo05,lo12}, quantum secret sharing \cite{Hillery99,Karlsson99,Gottesman00,Guo03,Xiao04,Zhang05}, Quantum Secure Direct Communication (QSDC) \cite{Long02,deng03,deng04,wang05,wan05,Wang06,jin06,Long07,xi07,lin08,Zhang17} etc., which have been  widely explored over the past 30 years.
	
	In classical  cryptography, sending a message from Alice to Bob always requires a key. In particular, one shared secret key is required for any symmetric key protocol and a pair of keys (one public key and one private key of the receiver Bob) is required for any asymmetric or public key protocol. Interestingly, in quantum domain there exist some protocols for secure message transmission that does not explicitly require any key. QSDC is one such protocol. The intuitive idea of QSDC was first proposed by G. L. Long and X. S. Liu in 2002~\cite{Long02}. In 2003, Deng et al. generalized the previous one and proposed another QSDC protocol~\cite{deng03}, where the sender (Bob) and receiver (Alice) first share two-particle entangled states (namely, one of the Bell state) and each of them takes one particle from each pair. After that, Bob encodes his state with one of the four unitary operations, which are called Pauli matrices \cite{Nielsen2002}, $I$, $\sigma_z$, $\sigma_x$, and $\sigma_{iy}$ to encode the information $00$, $01$, $10$, and $11$ respectively and sends it to Alice. Then Alice measures the two-particle state (one from Bob and another from her) in Bell basis to decode Bob's message. One of the famous QSDC protocol is Ping-Pong Protocol (PPP) \cite{ppp02}, where the receiver first prepares two qubit entangled states and ping the sender with one qubit. Then sender encodes her information by performing $I$ or $\sigma_z$ on that qubit and pong it to the receiver. Many other QSDC protocol have been analyzed in several works using different approaches \cite{deng04,wang05,wan05,Wang06,jin06,Long07,xi07,lin08,Zhang17}.
	
	Quantum Dialogue (QD) can be thought as a two way QSDC protocol. Nowadays it is a very important research topic in quantum cryptography. In QD, Alice and Bob can send messages to each other simultaneously in the same channel. Quantum dialogue was first proposed by BA Nguyen in 2004 \cite{nay04}. Nguyen first found out some drawback in the so-called PPP \cite{ppp02} and improved it. Then they extended the PPP to a QD protocol such that Alice and Bob can exchange their secret message directly. At the same time, Zhanjun Zhang also gave the idea of secure direct bidirectional communication \cite{zhang04}. In 2005 MAN Zhong-Xiao \textit{et al.} showed that the QD protocol proposed by Nguyen was insecure against intercept and resend attack strategy \cite{zhong05}. They modified the protocol in such a way that intercept and resend attack can be detected. After that, Yan XIA \textit{et al.} proposed a QD protocol using the GHZ state, which is also a modified version of Nguyen's protocol \cite{xia06}. In 2006, Ji Xin and Zhang Shou proposed a QD protocol based on single-photon \cite{xin06}. Recently various research work have been done in this area \cite{yan07,Yang07,Tan08,Dong08,Gao08,Gao10,Shi10,Yang13}.
	
	In 2017, A. Maitra proposed a Measurement Device Independent Quantum Dialogue (MDI-QD) protocol \cite{mai17}. In that protocol, there are two legitimate parties, namely Alice and Bob, who want to communicate simultaneously. There is an untrusted third party (UTP), who helps them to communicate. In the MDI-QD model, this UTP may itself act as an eavesdropper. First, Alice and Bob share a key using BB84 QKD \cite{bb84}. Then they prepare qubits corresponding to their messages and the shared key. They send their qubits the UTP. After receiving the qubits from Alice and Bob, the UTP measures the qubits and declares the results. From the measurement results, Alice and Bob guess the messages of each other. They discard almost half of the qubits to prevent information leakage and prove that their scheme is secure under this adversarial model. 
	
	\subsection*{Our Contributions}
	In this paper, we first revisit the MDI-QD protocol of A. Maitra \cite{mai17} in Section~\ref{sec2}. Then in Section~\ref{sec3} we propose two modifications of MDI-QD. We reduce the number of discarded qubits to almost the half of their count. In addition, we make use of some of their discarded qubits to communicate securely. In our two protocols we use two different techniques to choose the discarded qubits. For this, Alice and Bob generate some sequences depending on the key and the measurement results. Based on the sequences' terms, they decide which measurement results to keep. Details are given in Algorithm~\ref{algo:1qd} and Algorithm~\ref{algo:2qd}. For better understanding we give two examples of our protocols in Section~\ref{sec3.3}. We also discuss about the difference between the two protocols in Section~\ref{sec3.6}. Section~\ref{sec4} concludes our results.

	\subsection*{Notations}
	Throughout the paper we use some notations and we describe those common notations here.
	
	\begin{itemize}[label=$\bullet$]
		\item $Z$ basis $=\{\ket{0},\ket{1}\}$ basis;
		\item $\ket{+}=\frac{1}{\sqrt{2}}(\ket{0}+ \ket{1})$, $\ket{-}=\frac{1}{\sqrt{2}}(\ket{0}- \ket{1})$;
		\item $X$ basis $=\{\ket{+},\ket{-}\}$ basis;
		\item $I=\ket{0}\bra{0}+\ket{1}\bra{1}$;
		\item $\sigma_x=\ket{1}\bra{0}+\ket{0}\bra{1}$;
		\item $i\sigma_y=\ket{0}\bra{1}-\ket{1}\bra{0}$;
		\item $\sigma_z=\ket{0}\bra{0}-\ket{1}\bra{1}$;
		\item $\ket{\Phi^{+}}=\frac{1}{\sqrt{2}}(\ket{00}+ \ket{11})$, $\ket{\Phi^{-}}=\frac{1}{\sqrt{2}}(\ket{00}- \ket{11})$;
		\item $\ket{\Psi^{+}}=\frac{1}{\sqrt{2}}(\ket{01}+ \ket{10})$, $\ket{\Psi^{-}}=\frac{1}{\sqrt{2}}(\ket{01}- \ket{10})$;
		\item Bell basis $=\{\ket{\Phi^{+}},\ket{\Phi^{-}},\ket{\Psi^{+}},\ket{\Psi^{-}}\}$ basis;
		\item ${\{S[i]\}}_{i=1}^n=S$ is a finite sequence of length $n$;
		\item $S[i]=i$-th element of $S$ ;
		\item $\bar{b}$ = bit complement of $b$;
		\item $a \oplus b=$ $a ~ XOR ~b$
		\item $\Pr(A)=$ Probability of occurrence of an event $A$;
		\item $\Pr(A|B)=$ Probability of occurrence of an event $A$ given that the event $B$ has already occurred.
		
	\end{itemize}
	\section{Revisiting the MDI-QD Protocol of A. Maitra \cite{mai17}} \label{sec2}
	In this section we revisit the MDI-QD protocol proposed in \cite{mai17}. They composed two different protocols (BB84 \cite{bb84} and a modified version of measurement device independent quantum key distribution \cite{lo12}) to propose their protocol. There are two parts in their protocol. In the first part, two legitimate parties Alice and Bob perform BB84 QKD \cite{bb84} to generate a shared key $k$ between themselves. In the second part, they prepare their qubits corresponding to their message with the help of $k$. The encoding procedure is given in Algorithm~\ref{algo:1}.
	
	\begin{algorithm}
		\setlength{\textfloatsep}{0.05cm}
		\setlength{\floatsep}{0.05cm}
		Let the key be $k=k_1k_2\ldots k_n$, Alice's message be $a=a_1a_2\ldots a_n$ and Bob's message be $b=b_1b_2\ldots b_n$.
		
		Then for $1\leq i \leq n$, Alice and Bob prepare their qubits according to the following strategy:
		\begin{enumerate}
			\item if $a_i$ ($b_i $)$ = 0$ and $k_i = 0$, prepares $\ket{0}$.
			\item  if $a_i$ ($b_i $)$ = 1$ and $k_i = 0$, prepares $\ket{1}$.
			\item if $a_i$ ($b_i $)$ = 0$ and $k_i = 1$, prepares $\ket{+}$.
			\item  if $a_i$ ($b_i $)$ = 1$ and $k_i =1$, prepares $\ket{-}$.
		\end{enumerate}				
		\caption{Algorithm for encoding}
		\label{algo:1}
		\setlength{\textfloatsep}{0.05cm}
		\setlength{\floatsep}{0.05cm}
	\end{algorithm}
	
	Alice and Bob send their qubits to an UTP (Eve). Then UTP measures the two qubit states in Bell basis (i.e., $\{\ket{\Phi^{+}},\ket{\Phi^{-}},\ket{\Psi^{+}},\ket{\Psi^{-}}\}$ basis) and announces the result. From the result Alice (Bob) decodes the message of Bob (Alice) (see Table~\ref{table_qd}).
	
	\begin{table}[!htbp]
		\centering
		\caption{Different cases in MDI QSDC.}
		\begin{tabular}{|l|l|l|l|l|l|l|l|}
			\toprule
			\multicolumn{2}{|l|}{Message Bits of} & \multicolumn{2}{|l|}{Prepared qubits of} &  \multicolumn{4}{|l|}{Probability (Eve’s end)} \\
			\midrule
			{} Alice  & Bob &  Alice   &  Bob  & $\ket{\phi^+}$ & $\ket{\phi^-}$ & $\ket{\psi^+}$ & $\ket{\psi^-}$ \\
			\hline
			$0$ & $0$  & $\ket{0}$  &  $\ket{0}$ & $1/2$ & $1/2$  & $0$ & $0$\\
			$0$ & $1$  & $\ket{0}$  &  $\ket{1}$ &  $0$ & $0$ & $1/2$ & $1/2$\\
			$1$ & $0$  & $\ket{1}$  &  $\ket{0}$ &  $0$ & $0$ & $1/2$ & $1/2$\\
			$1$ & $1$  & $\ket{1}$  &  $\ket{1}$ & $1/2$ & $1/2$  & $0$ & $0$\\
			\hline
			$0$ & $0$ & $\ket{+}$  &  $\ket{+}$ & $1/2$ & $0$  & $1/2$ & $0$\\
			$0$ & $1$ & $\ket{+}$  &  $\ket{-}$ & $0$ & $1/2$  & $0$ & $1/2$\\
			$1$ & $0$ & $\ket{-}$  &  $\ket{+}$ & $0$ & $1/2$  & $0$ & $1/2$\\
			$1$ & $1$ & $\ket{-}$  &  $\ket{-}$ & $1/2$ & $0$  & $1/2$ & $0$\\
			\bottomrule
		\end{tabular}
		\label{table_qd}
	\end{table}
	
	It is clear from Table~\ref{table_qd} that, 
	\begin{itemize}
		\item if the prepared qubit of Alice is $\ket{0}$($\ket{1})$, then Alice guesses message bit of Bob with probability $1$ as follows:
		\begin{equation*}
		\text{Measurement result} =
		\begin{cases}
		\ket{\phi^+}$ or $\ket{\phi^-} \Rightarrow &\text{message bit of Bob is $0$ ($1$)}\\
		\ket{\psi^+}$ or $\ket{\psi^-} \Rightarrow &\text{message bit of Bob is $1$ ($0$)}
		\end{cases}
		\end{equation*}
		
		\item if the prepared qubit of Alice is $\ket{+}$($\ket{-})$, then Alice guesses message bit of Bob with probability $1$ as follows:
		\begin{equation*}
		\text{Measurement result} =
		\begin{cases}
		\ket{\phi^+}$ or $\ket{\psi^+} \Rightarrow &\text{message bit of Bob is $0$ ($1$)}\\
		\ket{\phi^-}$ or $\ket{\psi^-} \Rightarrow &\text{message bit of Bob is $1$ ($0$)}
		\end{cases}
		\end{equation*}
		
	\end{itemize}
	Similarly Bob can guess the communicated bit of Alice. Hence both can exchange their message simultaneously.
	
	Now we can see from Table~\ref{table_qd}, if the measurement result is $\ket{\phi^+}$ or $\ket{\psi^-}$ then Eve knows the XOR of the communicated bits between Alice and Bob.
	In that case Eve has $1$ bit information. To avoid the information leakage, Alice and Bob discard the measurement result when it is $\ket{\phi^+}$ or $\ket{\psi^-}$.
	
	After that, Alice and Bob estimate the error between the channel. If the UTP cheats, that can also be detected from this checking. If the error lies between a tolerable range they continue the protocol, else they abort. 
	
	\section{Efficient Measurement Device Independent Quantum Dialogue Protocols} \label{sec3}
	In the previous section, we discussed the MDI-QD protocol given in \cite{mai17}. Here we propose two efficient MDI-QD protocols which are modifications of \cite{mai17}. In our protocols, after the key generation step as \cite{mai17}, let the shared key between two legitimate parties Alice and Bob be $k=k_1k_2\ldots k_n$. They calculate the bit $c=\oplus k_i$, $1\leq i \leq n$. Then both of our protocols are the same as \cite{mai17} up-to the step where the UTP announces the measurement results. In the next step, Alice and Bob estimate the error in the channel (process is also same as \cite{mai17}). If the estimated error lies between a tolerable range they continue the protocol, else they abort. In the protocol of \cite{mai17}, Alice and Bob discard almost half of the measurement results. We reduce the number of discarded measurement results by generating some sequences and computing some functions of the sequences.
	
	\subsection{Our First Efficient Measurement Device Independent Quantum Dialogue Protocol} 
	After the error estimation phase, let the number of remaining measurement results be $n'$, Alice and Bob make a finite sequence $\{M[i]\}_{i=1}^{n'}$ containing the measurement results. i.e., $M[i]$ is the $i$-th measurement result announced by the UTP, for $1\leqslant i \leqslant n'$ and $M[i] \in \{\ket{\phi^+},\ket{\phi^-},\ket{\psi^+},\ket{\psi^-}\}$.
	They keep all the measurement results $M[i]$s where $M[i] \in \{\ket{\phi^-},\ket{\psi^+}\}$. 
	Among the remaining measurement results, they choose some of them to keep and discard the others. For $1\leq i \leq n'$, if $M[i] \in \{\ket{\phi^+},\ket{\psi^-}\}$ and $k_i=c$, then Alice and Bob keep that $M[i]$. Else they discard that $M[i]$. Using Table~\ref{tab:Alice' guess about $b_i$ } and Table~\ref{tab: Bob's guess about $a_i$ }, they guess the message bit of each other corresponding to all the measurement results $M[i]$ which they kept. Details are given in Algorithm~\ref{algo:1qd}.
	
	\begin{algorithm}
		\setlength{\textfloatsep}{0.05cm}
		\setlength{\floatsep}{0.05cm}
		\begin{enumerate}
			\item Alice and Bob share a $n$-bit key stream ($k=k_1k_2\ldots k_n$) between themselves using BB84 protocol.
			
			\item They calculate $c=\oplus k_i$, $1\leq i \leq n$.
			
			\item Let $n$-bit message of Alice and Bob be  $a=a_1a_2\ldots a_n$ and $b=b_1b_2\ldots b_n$ respectively.
			
			\item For $1\leq i \leq n$, Alice (Bob) prepares the qubits $Q_A~(Q_B)$ at her (his) end according to the following strategy:
			\begin{enumerate}
				\item if $a_i$ ($b_i $)$ = 0$ and $k_i = 0$, set ${Q_A}_i~({Q_B}_i)=\ket{0}$;
				\item if $a_i$ ($b_i $)$ = 1$ and $k_i = 0$, set ${Q_A}_i~({Q_B}_i)=\ket{1}$;
				\item if $a_i$ ($b_i $)$ = 0$ and $k_i = 1$, set ${Q_A}_i~({Q_B}_i)=\ket{+}$;
				\item if $a_i$ ($b_i $)$ = 1$ and $k_i =1$, set ${Q_A}_i~({Q_B}_i)=\ket{-}$.
			\end{enumerate}
			
			\item Alice (Bob) sends $Q_A~(Q_B)$ to the third party (TP). 
			
			\item For $1\leq i \leq n$, the UTP measures the two qubits ${Q_A}_i$ and ${Q_B}_i$ in Bell basis and announces the result.
			
			\item Alice and Bob make a finite sequence $\{M[i]\}_{i=1}^n$ containing the measurement results, i.e., for $1\leqslant i \leqslant n$, $M[i]$ is the $i$-th measurement result announced by the UTP, where $M[i] \in \{\ket{\phi^+},\ket{\phi^-},\ket{\psi^+},\ket{\psi^-}\}$.
			
			\item They randomly choose $\gamma n$ number of measurement results $M[i]$ from the sequence $\{M[i]\}_{i=1}^n$ to estimate the error, where $\gamma <1$ is a small fraction.
			
			\item Alice and Bob guess the message bit of other, corresponding to their chosen $\gamma n$ number of measurement results using Table~\ref{tab:Alice' guess about $b_i$ } and Table~\ref{tab: Bob's guess about $a_i$ }. 
			
			\item They reveal their respective guesses for these rounds.
			
			\item If estimated error is greater than some predefined threshold value, then they abort. Else continue and goto next step.
			
			\item Their remaining sequence of measurement results is relabeled as $\{M[i]\}_{i=1}^{n'}$, where $n'=(1-\gamma) n$.
			
			\item They update their $n$-bit key to an $n'$-bit key by discarding $\gamma n$ number of key bits corresponding to above $\gamma n$ rounds. The updated key is relabeled as $k=k_1k_2\ldots k_{n'}$.
			
			\item They generate a finite sequence $\{X[i]\}_{i=1}^{n'}$ such that 
			\begin{equation*}
			X[i]=
			\begin{cases}
			1, & \text{if $M_i=\ket{\phi^-}$ or $\ket{\psi^+}$; }\\
			0, & \text{otherwise.}
			\end{cases}
			\end{equation*}
			\item Then they generate another finite sequence $\{Y[i]\}_{i=1}^{n'}$ such that 
			\begin{equation*}
			Y[i]=
			\begin{cases}
			0, & \text{if $X[i]=1$; }\\
			k_j, & \text{if $c=1$ and $X[i]$ is the $j$-th zero of the sequence $\{X[q]\}_{q=1}^{n'}$; }\\
			\bar{k_j}, & \text{if $c=0$ and $X[i]$ is the $j$-th zero of the sequence $\{X[q]\}_{q=1}^{n'}$. }
			\end{cases}
			\end{equation*}
			\item For $1\leq i \leq n' $:
			\begin{itemize}
				\item if $X[i]\oplus Y[i]=1$, then Alice and Bob consider the $i$-th measurement result $M[i]$ and guess others message bit using Table~\ref{tab:Alice' guess about $b_i$ } and Table~\ref{tab: Bob's guess about $a_i$ }.
				\item Else they discard $M[i]$.
			\end{itemize}
			
		\end{enumerate}
		\caption{First proposed protocol for efficient MDI-QD}
		\label{algo:1qd}
		\setlength{\textfloatsep}{0.05cm}
		\setlength{\floatsep}{0.05cm}
	\end{algorithm}
	
	{
		\begin{algorithm}
			\setlength{\textfloatsep}{0.05cm}
			\setlength{\floatsep}{0.05cm}
			\begin{small}
				\begin{enumerate}
					
					\item Alice and Bob share a $n$-bit key stream ($k=k_1k_2\ldots k_n$) between themselves using BB84 protocol.
					
					\item They calculate $c=\oplus k_i$, $1\leq i \leq n$.
					
					\item Let $n$ bit message of Alice and Bob be  $a=a_1a_2\ldots a_n$ and $b=b_1b_2\ldots b_n$ respectively.
					
					\item For $1\leq i \leq n$, Alice (Bob) prepares the qubits $Q_A~(Q_B)$ at her (his) end according to the following strategy:
					\begin{enumerate}
						\item if $a_i$ ($b_i $)$ = 0$ and $k_i = 0$, set ${Q_A}_i~({Q_B}_i)=\ket{0}$;
						\item if $a_i$ ($b_i $)$ = 1$ and $k_i = 0$, set ${Q_A}_i~({Q_B}_i)=\ket{1}$;
						\item if $a_i$ ($b_i $)$ = 0$ and $k_i = 1$, set ${Q_A}_i~({Q_B}_i)=\ket{+}$;
						\item  if $a_i$ ($b_i $)$ = 1$ and $k_i =1$, set ${Q_A}_i~({Q_B}_i)=\ket{-}$.
					\end{enumerate}
					
					\item Alice (Bob) sends $Q_A~(Q_B)$ to the third party (TP). 
					
					\item For $1\leq i \leq n$, the UTP measures the two qubits ${Q_A}_i$ and ${Q_B}_i$ in Bell basis and announces the result.
					
					\item Alice and Bob make a finite sequence $\{M[i]\}_{i=1}^n$ containing the measurement results, i.e., for $1\leqslant i \leqslant n$, $M[i]$ is the $i$-th measurement result announced by the UTP, where $M[i] \in \{\ket{\phi^+},\ket{\phi^-},\ket{\psi^+},\ket{\psi^-}\}$.
					
					\item They randomly choose $\gamma n$ number of measurement results $M[i]$ from the sequence $\{M[i]\}_{i=1}^n$ to estimate the error, where $\gamma <1$ is a small fraction.
					
					\item Alice and Bob guess the message bit of other, corresponding to their chosen $\gamma n$ number of measurement results using Table~\ref{tab:Alice' guess about $b_i$ } and Table~\ref{tab: Bob's guess about $a_i$ }. 
					
					\item They reveal their respective guesses for these rounds.
					
					\item If estimated error is greater than some predefined threshold value, then they abort. Else continue and goto next step.
					
					\item Their remaining sequence of measurement results is relabeled as $\{M[i]\}_{i=1}^{n'}$, where $n'=(1-\gamma)n$.
					
					\item They update their $n$-bit key to an $n'$-bit key by discarding $\gamma n$ number of key bits corresponding to above $\gamma n$ rounds. The updated key is relabeled as $k=k_1k_2\ldots k_{n'}$.
					
					\item They generate a finite sequence $\{X[i]\}_{i=1}^{n'}$ such that 
					\begin{equation*}
					X[i]=
					\begin{cases}
					1, & \text{if $M_i=\ket{\phi^-}$ or $\ket{\psi^+}$; }\\
					0, & \text{otherwise.}
					\end{cases}
					\end{equation*}
					
					\item Then they generate another two finite sequence $\{Y[i]\}_{i=1}^{n'}$ and $\{Z[i]\}_{i=1}^{n'}$  such that 
					\begin{equation*}
					Y[i]=
					\begin{cases}
					0, & \text{if $X[i]=1$; }\\
					k_j, & \text{if $c=1$ and $X[i]$ is the $j$-th zero of the sequence $\{X[q]\}_{q=1}^{n'}$; }\\
					\bar{k_j}, & \text{if $c=0$ and $X[i]$ is the $j$-th zero of the sequence $\{X[q]\}_{q=1}^{n'}$. }
					\end{cases}
					\end{equation*}
					
					\begin{equation*}
					Z[i]=
					\begin{cases}
					0, & \text{if $X[i]=1$; }\\
					k_j, & \text{if $c=0$ and $X[i]$ is the $j$-th zero of the sequence $\{X[q]\}_{q=1}^{n'}$; }\\
					\bar{k_j}, & \text{if $c=1$ and $X[i]$ is the $j$-th zero of the sequence $\{X[q]\}_{q=1}^{n'}$. }
					\end{cases}
					\end{equation*}
					
					\item For Alice's message ($1\leq i \leq n' $):
					\begin{itemize}
						\item if $X[i]\oplus Y[i]=1$, then Alice and Bob consider the $i$-th measurement result $M[i]$. Bob guesses Alice's message bit $a_i$ using Table~\ref{tab: Bob's guess about $a_i$ }.
						\item Else they discard $M[i]$.
					\end{itemize}
					
					\item For Bob's message ($1\leq i \leq n' $):
					\begin{itemize}
						\item if $X[i]\oplus Z[i]=1$, then Alice and Bob consider the $i$-th measurement result $M[i]$. Alice guesses Bob's message bit $b_i$ using Table~\ref{tab:Alice' guess about $b_i$ }
						\item Else they discard $M[i]$.
					\end{itemize}
					
				\end{enumerate}
			\end{small}
			\caption{Second proposed protocol for efficient MDI-QD}
			\label{algo:2qd}
			\setlength{\textfloatsep}{0.05cm}
			\setlength{\floatsep}{0.05cm}
		\end{algorithm}
	}	
	
	\subsection{Our Second Efficient Measurement Device Independent Quantum Dialogue Protocol} 
	After the error estimation phase, let the number of remaining measurement results be $n'$, Alice and Bob make a finite sequence $\{M[i]\}_{i=1}^{n'}$ containing the measurement results. i.e., $M[i]$ is the $i$-th measurement result announced by the UTP, for $1\leqslant i \leqslant n'$ and $M[i] \in \{\ket{\phi^+},\ket{\phi^-},\ket{\psi^+},\ket{\psi^-}\}$.
	They keep all the measurement results $M[i]$s where $M[i] \in \{\ket{\phi^-},\ket{\psi^+}\}$.
	Among the remaining measurement results, they choose some to keep and discard other. 
	
	To choose the measurement results for Alice's message, they will do the following:\\
	for $1\leq i \leq n'$, if $M[i] \in \{\ket{\phi^+},\ket{\psi^-}\}$ and $k_i=c$, then Alice and Bob keep that $M[i]$. Else they discard that $M[i]$. Using Table~\ref{tab: Bob's guess about $a_i$ }, Bob guesses the message bit of Alice corresponding to all the measurement results $M[i]$ which they kept.
	
	To choose the measurement results for Bob's message, they will do the following:\\
	for $1\leq i \leq n'$, if $M[i] \in \{\ket{\phi^+},\ket{\psi^-}\}$ and $k_i=\bar{c}$, then Alice and Bob keep that $M[i]$. Else they discard that $M[i]$. Using Table~\ref{tab:Alice' guess about $b_i$ }, Alice guesses the message bit of Bob corresponding to all the measurement results $M[i]$ which they kept. In this case the length of final messages of Alice and Bob may differ.
	Details are given in Algorithm~\ref{algo:2qd}.

	\subsection{Examples of Quantum Dialogue using our proposed Protocols} \label{sec3.3}
	Let us take an example to understand our protocols more clearly. Here we skip the error estimation phase. 
	
	\subsubsection{Quantum Dialogue Protocol using Algorithm~\ref{algo:1qd}} 
	
	\begin{enumerate}
		\item Let $k=10011101101001010010$ be the shared key between Alice and Bob, then $c=\oplus k_i= 0$.
		\item Let Alice's message be $a=10110100111010110011$,
		\item Let Bob's message be $b=01101000101001101011$.
		\item Alice's encrypted message\\ $Q_A=\ket{-}\ket{0}\ket{1}\ket{-}\ket{+}\ket{-}\ket{0}\ket{+}\ket{-}\ket{1}\ket{-}\ket{0}\ket{1}\ket{+}\ket{1}\ket{-}\ket{0}\ket{0}\ket{-}\ket{1}$.
		\item Bob's encrypted message\\ $Q_B=\ket{+}\ket{1}\ket{1}\ket{+}\ket{-}\ket{+}\ket{0}\ket{+}\ket{-}\ket{0}\ket{-}\ket{0}\ket{0}\ket{-}\ket{1}\ket{+}\ket{1}\ket{0}\ket{-}\ket{1}$.
		\item Alice and Bob send their respective sequences of qubits $Q_A$ and $Q_B$ to the UTP and the UTP measures the two qubits (one from Alice and one from Bob) in Bell basis and announces the results.
		\item Let $M$ be the sequence\\ $\ket{\phi^-},\ket{\psi^+},\ket{\phi^+},\ket{\psi^-},\ket{\psi^-},\ket{\phi^-},\ket{\phi^-},\ket{\phi^+},\ket{\psi^+},\ket{\psi^-},\ket{\phi^+},\ket{\phi^+},\ket{\psi^-},$\\
		$\ket{\phi^-},\ket{\phi^-},\ket{\phi^-},\ket{\psi^+},\ket{\phi^-},\ket{\phi^+},\ket{\phi^-}$
		\item $X$ is the sequence $1,1,0,0,0,1,1,0,1,0,0,0,0,1,1,1,1,1,0,1$. 
		\item $Y$ is the sequence $0,0,0,1,1,0,0,0,0,0,0,1,0,0,0,0,0,0,0,0$.
		\item Then $X \oplus Y$ is the sequence $1,1,0,1,1,1,1,0,1,0,0,1,0,1,1,1,1,1,0,1$.
		\item Alice and Bob consider the $i$-th message bit pair $(a_i,b_i)$ if $X[i] \oplus Y[i]=1$. That is, they consider ${a'}=10101010011001$ as Alice's message and ${b'}=01010010110101$ as Bob's message.
	\end{enumerate}
	
	\subsubsection{Quantum Dialogue Protocol using Algorithm~\ref{algo:2qd}} 
	
	\begin{enumerate}
		\item Let $k=10011101101001010010$ be the shared key between Alice and Bob, then $c=\oplus k_i= 0$.
		\item Let Alice's message be $a=10110100111010110011$,
		\item Let Bob's message be $b=01101000101001101011$.
		\item Alice's encrypted message\\ $Q_A=\ket{-}\ket{0}\ket{1}\ket{-}\ket{+}\ket{-}\ket{0}\ket{+}\ket{-}\ket{1}\ket{-}\ket{0}\ket{1}\ket{+}\ket{1}\ket{-}\ket{0}\ket{0}\ket{-}\ket{1}$.
		\item Bob's encrypted message\\ $Q_B=\ket{+}\ket{1}\ket{1}\ket{+}\ket{-}\ket{+}\ket{0}\ket{+}\ket{-}\ket{0}\ket{-}\ket{0}\ket{0}\ket{-}\ket{1}\ket{+}\ket{1}\ket{0}\ket{-}\ket{1}$.
		\item Alice and Bob send their respective sequences of qubits $Q_A$ and $Q_B$ to the UTP and the UTP measures the two qubits (one from Alice and one from Bob) in Bell basis and announces the results.
		\item Let $M$ be the sequence\\ $\ket{\phi^-},\ket{\psi^+},\ket{\phi^+},\ket{\psi^-},\ket{\psi^-},\ket{\phi^-},\ket{\phi^-},\ket{\phi^+},\ket{\psi^+},\ket{\psi^-},\ket{\phi^+},\ket{\phi^+},\ket{\psi^-},$\\
		$\ket{\phi^-},\ket{\phi^-},\ket{\phi^-},\ket{\psi^+},\ket{\phi^-},\ket{\phi^+},\ket{\phi^-}$
		\item $X$ is the sequence $1,1,0,0,0,1,1,0,1,0,0,0,0,1,1,1,1,1,0,1$. 
		\item $Y$ is the sequence $0,0,0,1,1,0,0,0,0,0,0,1,0,0,0,0,0,0,0,0$.
		\item $Z$ is the sequence $0,0,1,0,0,0,0,1,0,1,1,0,1,0,0,0,0,0,1,0$.
		\item Then $X \oplus Y$ is the sequence $1,1,0,1,1,1,1,0,1,0,0,1,0,1,1,1,1,1,0,1$ and
		\item $X \oplus Z$ is the sequence $1,1,1,0,0,1,1,1,1,1,1,0,1,1,1,1,1,1,1,1$.
		\item For Alice's message, Alice and Bob consider the $i$-th ($1\leq i \leq 20$) measurement result $M[i]$ only when $X[i]\oplus Y[i]=1$ and discard other cases. That is, they consider ${a'}=10101010011001$ as Alice's message. 
		\item For Bob's message, Alice and Bob consider the $i$-th ($1\leq i \leq 20$) measurement result $M[i]$ only when $X[i]\oplus Z[i]=1$ and discard other cases. That is, they consider ${b'}=01100010101101011$ as Bob's message.
	\end{enumerate}
	
	\subsection{Correctness of Our Proposed Protocols}
	In our proposed protocols, Alice and Bob first prepare qubits corresponding to their messages and shared key and then send those qubits to the third party (TP). After that, the UTP measures each two qubit state (one from Alice and one from Bob) in Bell basis and announces the result. Now, there may arise four cases and from help of Table~\ref{table_qd} we can say the followings:
	\begin{itemize}
		\item if the prepared qubit of Alice is $\ket{0}$($\ket{1})$, then Alice guesses message bit of Bob with probability $1$ as follows:
		\begin{equation*}
		\text{Measurement result} =
		\begin{cases}
		\ket{\phi^+}$ or $\ket{\phi^-} \Rightarrow &\text{message bit of Bob is $0$ ($1$)}\\
		\ket{\psi^+}$ or $\ket{\psi^-} \Rightarrow &\text{message bit of Bob is $1$ ($0$)}
		\end{cases}
		\end{equation*}
		
		\item if the prepared qubit of Alice is $\ket{+}$($\ket{-})$, then Alice guesses message bit of Bob with probability $1$ as follows:
		\begin{equation*}
		\text{Measurement result} =
		\begin{cases}
		\ket{\phi^+}$ or $\ket{\psi^+} \Rightarrow &\text{message bit of Bob is $0$ ($1$)}\\
		\ket{\phi^-}$ or $\ket{\psi^-} \Rightarrow &\text{message bit of Bob is $1$ ($0$)}
		\end{cases}
		\end{equation*}
		
	\end{itemize}
	
	From the above knowledge, we construct Table~\ref{tab:Alice' guess about $b_i$ }, which contents the information of Alice's guess about Bob's message for different cases.
	\begin{table}[h]
		\centering
		\caption{Alice's guess about Bob's message for different cases}
		\begin{tabular}{|l|l|l|l|l|l|l|l|l|}
			\toprule
			Key  & Alice's & Alice's &  \multicolumn{4}{|l|}{Alice's guess about $b_i$ when $M[i]$}\\
			bit $k_i$ & bit $a_i$ & qubit & $\ket{\phi^+}$ &$\ket{\phi^-}$ &$\ket{\psi^+}$ &$\ket{\psi^-}$\\
			\midrule
			0 & 0 & $\ket{0}$  & 0 & 0 & 1 & 1 \\
			
			0 & 1 & $\ket{1}$  & 1 & 1 & 0 & 0 \\
			
			1 & 0 & $\ket{+}$  & 0 & 1 & 0 & 1 \\
			
			1 & 1 & $\ket{-}$  & 1 & 0 & 1 & 0 \\
			\bottomrule
		\end{tabular}

		\label{tab:Alice' guess about $b_i$ }
	\end{table}

	Similar thing happens for Bob too. So we construct Table~\ref{tab: Bob's guess about $a_i$ }, which contents the information of Bob's guess about Alice's message for different cases.
	\begin{table}[h]
		\centering
		\caption{Bob's guess about Alice's message for different cases}
		\begin{tabular}{|l|l|l|l|l|l|l|l|l|}
			\toprule
			Key  & Bob's & Bob's &  \multicolumn{4}{|l|}{Bob's guess about $a_i$ when $M[i]$}\\
			bit $k_i$ & bit $b_i$ & qubit & $\ket{\phi^+}$ &$\ket{\phi^-}$ &$\ket{\psi^+}$ &$\ket{\psi^-}$\\
			\midrule
			0 & 0 & $\ket{0}$  & 0 & 0 & 1 & 1 \\
			
			0 & 1 & $\ket{1}$  & 1 & 1 & 0 & 0 \\
			
			1 & 0 & $\ket{+}$  & 0 & 1 & 0 & 1 \\
			
			1 & 1 & $\ket{-}$  & 1 & 0 & 1 & 0 \\
			\bottomrule
		\end{tabular}

		\label{tab: Bob's guess about $a_i$ }
	\end{table}
	
	From Table~\ref{tab:Alice' guess about $b_i$ } and Table~\ref{tab: Bob's guess about $a_i$ }, we see that for all cases Alice and Bob can conclude the communicated bit of the other party with probability $1$. That is, always they can guess the correct message bit of the other party with probability $1$. Hence both of our protocols are giving the correct results.

	\subsection{Security Analysis of Our Proposed Protocols}
	Both of our proposed protocols for Quantum Dialogue are modifications ofthe Quantum Dialogue protocol given in \cite{mai17}. In their protocol they have considered only the cases where the measurement results were $\ket{\phi^-}$ or $\ket{\psi^+}$ and discard the cases for $\ket{\phi^+}$ and $\ket{\psi^-}$. But in our protocols, we have used all the cases where the measurement results are $\ket{\phi^-}$, $\ket{\psi^+}$ and also some cases where the measurement results are $\ket{\phi^+}$, $\ket{\psi^-}$. We have done some classical computation to choose which results to take. Since in \cite{mai17}, the authors had done the security analysis of the protocol for the cases where the measurement results were $\ket{\phi^-}$ or $\ket{\psi^+}$, so it is sufficient for us to analyze the security of rest of the part of the protocols. 
	
	Before we proceed, let us first define the advantage of an adversary. It measures the success of an attack by an adversary on a cryptographic scheme. The advantage distinguishes the output of a cryptographic algorithm from that of a uniformly random source. If the advantage of an adversary for an algorithm is negligible, i.e., it is less than some predefined threshold value, then the algorithm is said to be secure. The word ``negligible" usually means ``within $O(2^{-p})$" where $p$ is a security parameter associated with the algorithm. 
	
		\textbf{Advantage}: For our purpose, the advantage of an adversary \textit{A} is the absolute value of the differences between the probabilities of the events $A_0$ and $A_1$, where  $A_0=$ Guessing a random message ``$m$" from the message space, and $A_1=$ Guessing the same message ``$m$" from the message space using our algorithm. That is, $Adv(A)=|\Pr(A_0)-\Pr(A_1)|$. 
	
	Our protocol is said to be secure if $Adv(A)< \epsilon $, where $\epsilon$ is the security parameter.
	
	We have an $n$ bit key $k=k_1k_2\ldots k_n$ and $c=\oplus k_i$, $1\leq i \leq n$. Alice's $n$ bit message is $a$ and Bob's $n$ bits message is $b$. Let there be $l$ number of zeros in the finite sequence $\{X[q]\}_{q=1}^n$. The UTP knows the value of $a_j \oplus b_j$ if $X[j]=0$ (when $X[j]=0$, the UTP knows that the communicated bits of Alice and Bob are same or different). 
	Let us consider the following.
	\begin{itemize}
		\item $k'=k'_1k'_2\ldots k'_l$, where $k'_i=k_j$ if $X[j]$ is $i$-th zero in the finite sequence $\{X[q]\}_{q=1}^n$.
		\item $e=l$ bit substring of $a$, where ${e}_i=a_j$, if $X[j]$ is the $i$-th zero of the sequence $\{X[q]\}_{q=1}^n$. 
		\item $f=l$ bit substring of $b$, where ${f}_i=b_j$, if $X[j]$ is the $i$-th zero of the sequence $\{X[q]\}_{q=1}^n$. 
		\item The UTP knows $e \oplus f$.
	\end{itemize}
	
	\subsubsection{Security Analysis of our first Proposed Protocol}
	\label{security analysis}
	In our first protocol, we keep the $i$-th ($1 \leqslant i \leqslant l$) message pair $({e}_i,{f}_i)$ if $k'_i=c$ and discard the others. Let $c_1=$ Number of cases where $k'_i=c$,  $1 \leqslant i \leqslant l$. 
	Let us define some events first.
	\begin{itemize}
		\item $E_0=$ Keeping the $i$-th message bit pair $({e}_i,{f}_i)$.
		\item $E_1=$ Knowing our new message pair.
		\item $E_2=$ Guessing a random message pair $(e,f)$ of length $c_1$.
	\end{itemize}
	
	So, $\Pr(E_0)=\frac{1}{2}$, $\Pr(({e}_i,{f}_i)|{e}_i \oplus {f}_i)=\frac{1}{2}$.
	
	Thus, $\Pr(E_1)=\left(\frac{1}{2}\right)^l\left(\frac{1}{2}\right)^{c_1}$. Again, $\Pr(E_2)=\left(\frac{1}{4}\right)^{c_1}$. 
	
	Now the expected value of $c_1=\frac{l}{2}$. Substituting this in the above expression, we get $\Pr(E_1) \approx \left(\frac{1}{2}\right)^{\frac{3l}{2}}$ and $\Pr(E_2) \approx \left(\frac{1}{4}\right)^{\frac{l}{2}}$. 
	
	Hence the advantage is, $Adv(A)=|\Pr(E_2)-\Pr(E_1)| \approx |{\left(\frac{1}{4}\right)^\frac{l}{2}-\left(\frac{1}{2}\right)^{\frac{3l}{2}}}|=\left(\frac{1}{2}\right)^{l}\left[1-\left(\frac{1}{2}\right)^{\frac{l}{2}}\right]$. 
	
	Now $Adv(A) < \epsilon$\\
	$ \Leftrightarrow \left( \frac{1}{2}\right)^{l}\left[1-\left(\frac{1}{2}\right)^{\frac{l}{2}}\right] < \epsilon$\\
	$ \Rightarrow \left(\frac{1}{2}\right)^{\frac{3l}{2}} \leqslant \left(\frac{1}{2}\right)^{l}\left[1-\left(\frac{1}{2}\right)^{\frac{l}{2}}\right] < \epsilon$ (assuming that  $\left(\frac{1}{2}\right)^{\frac{l}{2}} <  1-\left(\frac{1}{2}\right)^{\frac{l}{2}} \Leftrightarrow \left(\frac{1}{2}\right)^{{\frac{l}{2}}-1} < 1 \Leftrightarrow {{\frac{l}{2}}-1} > 0 \Leftrightarrow l>2 $.)\\ 
	$\Rightarrow \left(\frac{1}{2}\right)^{\frac{3l}{2}} < \epsilon \Leftrightarrow {-\frac{3l}{2}} < log(\epsilon) \Leftrightarrow l > \frac{2}{3} log(\frac{1}{\epsilon})$.
	
	So for a predefined security parameter $\epsilon$, if $l > $max$\{2, \frac{2}{3} log(\frac{1}{\epsilon})\}$, then $Adv(A)< \epsilon $, i.e., our protocol is secure. We can also adjust the value of $l$ by padding some random message bits.
	
	\subsubsection{Security Analysis of our second Proposed Protocol}
	In our second protocol, we keep the $i$-th bit of Alice's message ${e}_i$ if $k'_i=c$, the $i$-th bit of Bob's message ${f}_i$ if $k'_i=\bar{c}$, $1 \leqslant i \leqslant l$ and discard the rest.
	
	Let $c_1=$ Number of cases where $k'_i=c$,  $1 \leqslant i \leqslant l$.
	Let us define some events first.
	\begin{itemize}
		\item $E_0=$ Keeping $e_i$, the $i$-th message bit of Alice.
		\item $E_1=$ Keeping $f_i$, the $i$-th message bit of Bob.
		\item $E_2=$ Knowing Alice's and Bob's new message ${e}'$ and ${f}'$ respectively.
		\item $E_4=$ Guessing two random message $e$ and $f$ of length $c_1$ and $l-c_1$ respectively.
	\end{itemize}
	
	So, $\Pr(E_0)=\frac{1}{2}$ and $\Pr(E_1)=\frac{1}{2}$.
	
	Using the expectation of $c_1$ calculated earlier, we have $\Pr(E_3)=\left(\frac{1}{2}\right)^l\left(\frac{1}{2}\right)^{c_1}\left(\frac{1}{2}\right)^{l-c_1} \approx \left(\frac{1}{2}\right)^{2l}$. Again, $\Pr(E_4)=\left(\frac{1}{2}\right)^{c_1}\left(\frac{1}{2}\right)^{l-c_1} \approx \left(\frac{1}{2}\right)^{l}$. 
	
	Thus, the advantage of the UTP is,  $Adv(A)=|\Pr(E_4)-\Pr(E_3)| \approx |{\left(\frac{1}{2}\right)^l-\left(\frac{1}{2}\right)^{2l}}|=\left(\frac{1}{2}\right)^l\left[1-\left(\frac{1}{2}\right)^l\right]$.
	
	Now $Adv(A)< \epsilon $\\
	$\Leftrightarrow \left(\frac{1}{2}\right)^l\left[1-\left(\frac{1}{2}\right)^l\right] < \epsilon$\\
	$\Rightarrow \left(\frac{1}{2}\right)^{2l}< \left(\frac{1}{2}\right)^l\left[1-\left(\frac{1}{2}\right)^l\right] < \epsilon $  (assuming that  $\left(\frac{1}{2}\right)^{l} <  1-\left(\frac{1}{2}\right)^{l} \Leftrightarrow \left(\frac{1}{2}\right)^{{l}-1} < 1 \Leftrightarrow  l>1 $.)\\ 
	$\Rightarrow \left(\frac{1}{2}\right)^{2l}< \epsilon \Leftrightarrow {-2l} < log(\epsilon) \Leftrightarrow l > \frac{1}{2} log(\frac{1}{\epsilon}) $.
	
	So for a predefined security parameter $\epsilon$, if $l > $max$\{1, \frac{1}{2} log(\frac{1}{\epsilon})\}$, then $Adv(A)< \epsilon $, i.e., our protocol is secure. We can also adjust the value of $l$ by padding some random message bits.
	\subsection{Difference Between Our Two Protocols} \label{sec3.6}
	Both of our proposed protocols for quantum dialogue are modifications of the quantum dialogue protocol given in~\cite{mai17}. In these protocols, the UTP measures each two qubit state (one from Alice and one from Bob) in Bell basis and announces the result. Alice and Bob make a finite sequence $\{M[i]\}_{i=1}^n$ containing the measurement results. That is, $M[i]$ is the $i$-th ($1\leqslant i \leqslant n$) measurement result announced by the UTP and $M[i] \in \{\ket{\phi^+},\ket{\phi^-},\ket{\psi^+},\ket{\psi^-}\}$. After the error estimation phase, the remaining sequence of measurement results is relabeled as $\{M[i]\}_{i=1}^{n'}$. For $1\leqslant i \leqslant n$, if $M[i] \in \{\ket{\phi^-}, \ket{\psi^+}$\}, then we keep those results for both the protocols. But if $M[i] \in \{\ket{\phi^+}, \ket{\psi^-}\}$, then we use some technique to decide whether we keep those results or discard them.
	
	The basic difference between our two protocols is the technique of choosing $M[i]$ when $M[i]=\ket{\phi^+}$ or $\ket{\psi^-}$, $1\leq i \leq n' $. From our first protocol, we get a synchronized message pair of Alice and Bob. Here by synchronized message, we mean that if we keep the $i$-th message bit of Alice, then we also keep the $i$-th message bit of Bob. For this protocol, we consider the $i$-th message bit pair $(a_i,b_i)$, if $X[i]\oplus Y[i]=1$ holds ($1\leq i \leq n' $), where $\{X[i]\}_{i=1}^{n'}$ and $\{Y[i]\}_{i=1}^{n'}$ are defined in Algorithm~\ref{algo:1qd}.
	
	But for the second protocol, we do not get any synchronized message pair of Alice and Bob. In this protocol, if $M[i]=\ket{\phi^+}$ or $\ket{\psi^-}$, then, for some cases we keep the corresponding message bit of Alice and discard Bob's message bit, or the converse. For $1\leq i \leq n'$, the condition for keeping Alice's message bit $a_i$ is $X[i]\oplus Y[i]=1$, i.e., when $X[i]\oplus Y[i]=1$, we keep $a_i$ and discard $b_i$. Also for $1\leq i \leq n' $, the condition for keeping Bob's message bit $b_i$ is $X[i]\oplus Z[i]=1$, i.e., when $X[i]\oplus Z[i]=1$, we keep $b_i$ and discard $a_i$, where $\{X[i]\}_{i=1}^{n'}$, $\{Y[i]\}_{i=1}^{n'}$ and $\{Z[i]\}_{i=1}^{n'}$ are defined in Algorithm~\ref{algo:2qd}. 
	So for each $i$, we are keeping $a_i$ or $b_i$ or both.
	
	The performance of our second protocol is better, when $c_1< \frac{l}{2}$ (these are defined in Section~\ref{security analysis}). In that case, we can keep more message bits using our second protocol than the first one. One may note that synchronization is not an issue if only message transmission is considered. But if Alice and Bob use the synchronized messages to define something else, then our second protocol cannot be used (as the length of their final message may differ from each other). For this case, they have to use our first protocol.
	
	\section{Conclusion} \label{sec4}
	In this paper, we have proposed two protocols for quantum dialogue such that two legitimate parties Alice and Bob can securely communicate their messages simultaneously. Both of our proposed protocols are modifications of MDI-QD protocol given in \cite{mai17}. In their protocol they have used only half of the qubits. But in our protocols we have used almost three fourth of the qubits. So our protocols are more efficient than the previous one in terms of number of qubits. We have showed that our QD protocols are secure as advantages of adversary are negligible for both the cases. Also we have discussed about the difference between our two protocols.
	
	\section*{Acknowledgement}
	The first author would like to acknowledge Diptendu Chatterjee of Applied Statistics Unit, Indian Statistical Institute for comments that help improving the presentation of Section 3.5.

\end{document}